\documentstyle[amsfonts,aps]{revtex}

\begin{document}
\title{Scattering and intrinsic irreversibility}
\author{Roberto Laura}
\address{Departamento de F\'{\i}sica\\
FCEIA, Universidad Nacional de Rosario\\
Av. Pellegrini 250, 2000 Rosario, Argentina.\\
e-mail: laura@ifir.ifir.edu.ar}
\date{March 10th.,1997}
\maketitle

\begin{abstract}
The formalism of quantum systems with diagonal singularities is applied to
describe scattering processes. Well defined states are obtained for infinite
time, which are related to a ''weak form'' of intrinsic irreversibility.
Real and complex generalized spectral decompositions of the Liouville-Von
Neumann superoperator are computed. The physical meaning of ''Gamov states''
is discussed.
\end{abstract}

\section{Introduction.}

The search of a physical explanation for the evolution towards equilibrium
of quantum systems has a great interest for quantum statistical mechanics.
For many years a great number of papers were devoted to this problem.

The microscopic explanation of the approach to equilibrium was related to
the so called 'intrinsic irreversibility' of quantum systems. B.Misra,
I.Prigogine, M.Courbage \cite{6}\cite{7} pointed out the existence of a time
operator for the statistical description of classical and quantum systems.
The mean value of this operator is the 'age' of the system, which is a
growing function of time.

A. Bohm et al. \cite{8}\cite{9} related the intrinsic irreversibility to the
existence of generalized eigenvectors of the Hamiltonian with complex
eigenvalues, corresponding to poles of the analytic extension of the
scattering matrix.

Complex eigenvalues have been obtained by E. Sudarshan et al. \cite{10} by
analytic continuation in a generalized quantum mechanics.

The Friedrichs model, a prototype of decaying system describing the
interaction between a quantum oscillator and a scalar field, was extensively
analyzed in the literature for the one exited mode sector. It is an exactly
solvable model, in which the quantum oscillator decays to the ground state
for all initial conditions. E. Sudarshan et al. \cite{10} computed the
complex spectral decomposition. The spectral decomposition was also obtained
by T.Petrosky, I.Prigogine, S.Tasaki \cite{16} using subdynamic theory. The
spectral decomposition with complex eigenvalues was interpreted in terms of
Rigged Hilbert spaces by I.Antoniou, I.Prigogine in reference \cite{17} and
by I.Antoniou, S.Tasaki in reference \cite{11}.

When it is necessary to deal with systems with a huge number of particles,
the standard procedure is to start with $N$ particles in a box of volume $V$%
, making the limit $N\rightarrow \infty $, $V\rightarrow \infty $ with $%
\frac NV=c<\infty $ in the last step of the calculations. This method was
used in subdynamic theory (I.Antoniou, S.Tasaki \cite{11}, T.Petrosky,
I.Prigogine \cite{12}), where the collision operator, with complex
eigenvalues, is responsible for the evolution to statistical equilibrium.

It is not surprising that the time evolution of Friedrichs model can be
successfully described using the methods of non equilibrium statistical
mechanics that can be used, for example, to describe the approach to
statistical equilibrium of a quantum gas. In both cases the interaction
eliminates constants of motion. In Friedrichs model the discrete eigenvalue
disappear and in the gas the momentum of each particle is no more a constant
of motion when the interaction is present.

In this paper we want to discuss ''intrinsic irreversibility'' in connection
with pure scattering processes, where the total and the free Hamiltonian
have the same continuous spectrum. For this purpose, it is important to use
a formalism where ''final'' states ($t\rightarrow \infty $) are well defined.

For finite systems with continuous spectrum, the usual formalism of quantum
mechanics fails to give a description of the ''final'' states in terms of
wave functions or density operators. To overcame this difficulty we will use
in this paper the formalism developed by I. Antoniou et al. for quantum
systems with diagonal singularity \cite{13} \cite{14} \cite{15}. The quantum
states of this theory are ${\sl functionals}$ over the space of observables $%
{\cal O}$. Mathematically this means that the space ${\cal S}$ of states is
contained in ${\cal O}^{\times }$. Physically it means that the only thing
we can really observe and measure are the mean values of the observables $%
O\in {\cal O}$ in states $\rho \in {\cal S\subset O}^{\times }$: namely $%
\langle O\rangle _\rho =\rho [O]\equiv (\rho |O)$. This is the natural
generalization of the usual trace $Tr(\widehat{\rho }\widehat{O})$ which is
ill defined in systems with continuous spectrum. For finite quantum systems
with continuous spectrum, some observables (for example the Hamiltonian) are
represented by operators with diagonal singularities, and as they should
have well defined mean values, diagonal singularities also appears in the
states.

In section 2, the resolvent formalism including creation, destruction and
collision superoperators is obtained in general for quantum systems with
diagonal singularities.

In section 3, we apply this formalism to the scattering problem, showing
that the collision superoperator is zero, and computing the singular
''final'' state, and discuss its relation with a sort of ''weak intrinsic
irreversibility''.

In section 4, we compute the real and the complex spectral decompositions of
the time evolution with the help of the Lipmann-Schwinger vectors and its
analytic extensions. Complex eigenvalues appear related to the assumed
simple pole of the analytic extension of the density matrix. The physical
meaning of ''Gamov states'' is also discussed in this section.

\section{Resolvent formalism for generalized states.}

Let us consider the Liouville-Von Neumann equation for a state $\rho $%
\begin{equation}
i\frac d{dt}\rho ={\Bbb L}\rho  \label{2.1}
\end{equation}

The general solution, valid for $t>0$, can be written us 
\begin{equation}
\rho _t=\frac 1{2\pi i}\int\limits_\Gamma dz\,\exp (-izt)\frac 1{{\Bbb L}-z}%
\rho _o  \label{2.2}
\end{equation}
where $\Gamma $ is an horizontal line parallel to the real axis and located
in the upper half plane.

By defining two projectors ${\Bbb P}$ and ${\Bbb Q}$ acting on the states
and satisfying 
\begin{equation}
{\Bbb P}^2={\Bbb P},\qquad {\Bbb Q}^2={\Bbb Q},\qquad {\Bbb P+Q=I},
\label{2.3}
\end{equation}
(${\Bbb I}$ is the identity superoperator acting on states), it is possible
to decompose the resolvent as \cite{Grecos}\cite{Zwan} 
\begin{equation}
\frac 1{{\Bbb L}-z}=\left[ {\Bbb P+C}(z)\right] \frac 1{{\Bbb PLP+}\Psi (z)-z%
}\left[ {\Bbb P+D}(z)\right] +\frac 1{{\Bbb QLQ-}z}{\Bbb Q}.  \label{2.4}
\end{equation}

The superoperators $\Psi (z)$, ${\Bbb C}(z)$ and ${\Bbb D}(z)$ of the
previous expression are defined by: 
\begin{eqnarray}
\Psi (z) &=&{\Bbb P}\Psi (z){\Bbb P=}-{\Bbb PLQ}\frac 1{{\Bbb QLQ-}z}{\Bbb %
QLP},  \nonumber \\
{\Bbb C}(z) &=&{\Bbb QC(}z{\Bbb )P=-}\frac 1{{\Bbb QLQ}-z}{\Bbb QLP},
\label{2.5} \\
{\Bbb D}(z) &=&{\Bbb PD}(z){\Bbb Q=-PLQ}\frac 1{{\Bbb QLQ-}z}.  \nonumber
\end{eqnarray}
They are called collision, creation and destruction operators.

As we wish to include generalized states in the formalism, it is useful to
consider that the states are represented by antilinear functionals acting on
the representation of the observables to give the mean value\cite{13}\cite
{14}\cite{15}. Therefore for any observable $O\in {\cal O}$ and state $\rho
\in {\cal O}^{\times }$ we have: 
\begin{equation}
\langle O\rangle _\rho =(\rho |O),\qquad (a\rho _1+b\rho _2|O)=a^{*}(\rho
_1|O)+b^{*}(\rho _2|O).  \label{2.6}
\end{equation}

As it is usual in quantum mechanics, the observables are represented by self
adjoint operators for which we expect real mean values: 
\begin{equation}
(\rho |O)=(\rho |O)^{*}.  \label{2.7}
\end{equation}

The identity operator can be written as the sum over the projections on all
the generalized pure states, i.e. $I=\sum_\alpha |\alpha \rangle \langle
\alpha |$. Therefore we should impose 
\begin{equation}
(\rho |I)=\sum_\alpha (\rho ||\alpha \rangle \langle \alpha |)=1,\qquad
(\rho ||\alpha \rangle \langle \alpha |)\geq 0.  \label{2.8}
\end{equation}
In the last expression $(\rho ||\alpha \rangle \langle \alpha |)$ is the
probability of the state $\rho $ to be the pure state $|\alpha \rangle $.
For continuous spectrum the sum in (\ref{2.8}) should be replaced by an
integral and $(\rho ||\alpha \rangle \langle \alpha |)$ is a density of
probability. Expression (\ref{2.8}) is the generalization to states
represented by functionals of the concept of trace.

In this approach it is necessary to reconsider the Liouville-Von Neumann
equation (\ref{2.1}), which can be applied to an arbitrary observable $O$,
i.e.: 
\[
(i\frac d{dt}\rho |O)=({\Bbb L}\rho |O). 
\]

We should give a meaning to the second term of the previous equation. The
superoperator ${\Bbb L}$ is defined by 
\begin{equation}
({\Bbb L}\rho |O)=(\rho |{\Bbb L}^{\dagger }O)=(\rho |[H,O]),  \label{2.9}
\end{equation}
where $H$ is the Hamiltonian of the quantum system.

From (\ref{2.2}) and the antilinearity of the state functionals we obtain 
\begin{equation}
(\rho _t|O)=\frac i{2\pi }\int\limits_\Gamma dz\,\exp (-izt)(\rho _o|\frac 1{%
{\Bbb L}^{\dagger }+z}|O).  \label{2.10}
\end{equation}

The resolvent $\frac 1{{\Bbb L}^{\dagger }+z}$ has the following
decomposition 
\begin{equation}
\frac 1{{\Bbb L}^{\dagger }+z}=\left[ {\Bbb P}^{\dagger }{\Bbb +D}^{\dagger
}(z)\right] \frac 1{{\Bbb P}^{\dagger }{\Bbb L}^{\dagger }{\Bbb P}^{\dagger }%
{\Bbb +}\Psi ^{\dagger }(z)+z}\left[ {\Bbb P}^{\dagger }{\Bbb +C}^{\dagger
}(z)\right] +{\Bbb Q}^{\dagger }\frac 1{{\Bbb Q}^{\dagger }{\Bbb L}^{\dagger
}{\Bbb Q}^{\dagger }{\Bbb +}z},  \label{2.11}
\end{equation}
where ${\Bbb P}^{\dagger }$ and ${\Bbb Q}^{\dagger }$ are defined by 
\begin{equation}
(\rho |{\Bbb P}^{\dagger }O)=({\Bbb P}\rho |O),\qquad (\rho |{\Bbb Q}%
^{\dagger }O)=({\Bbb Q}\rho |O),  \label{2.12}
\end{equation}
and 
\begin{eqnarray}
\Psi ^{\dagger }(z) &=&{\Bbb P}^{\dagger }\Psi ^{\dagger }(z){\Bbb P}%
^{\dagger }{\Bbb =}-{\Bbb P}^{\dagger }{\Bbb L}^{\dagger }{\Bbb Q}^{\dagger }%
\frac 1{{\Bbb Q}^{\dagger }{\Bbb L}^{\dagger }{\Bbb Q}^{\dagger }{\Bbb +}z}%
{\Bbb Q}^{\dagger }{\Bbb L}^{\dagger }{\Bbb P}^{\dagger },  \nonumber \\
{\Bbb C}^{\dagger }(z) &=&{\Bbb P}^{\dagger }{\Bbb C}^{\dagger }{\Bbb (}z%
{\Bbb )Q}^{\dagger }{\Bbb =-P}^{\dagger }{\Bbb L}^{\dagger }{\Bbb Q}%
^{\dagger }\frac 1{{\Bbb Q}^{\dagger }{\Bbb L}^{\dagger }{\Bbb Q}^{\dagger
}+z},  \label{2.13} \\
{\Bbb D}^{\dagger }(z) &=&{\Bbb Q}^{\dagger }{\Bbb D}^{\dagger }(z){\Bbb P}%
^{\dagger }{\Bbb =-}\frac 1{{\Bbb Q}^{\dagger }{\Bbb L}^{\dagger }{\Bbb Q}%
^{\dagger }{\Bbb +}z}{\Bbb Q}^{\dagger }{\Bbb L}^{\dagger }{\Bbb P}^{\dagger
}.  \nonumber
\end{eqnarray}

\section{Intrinsic irreversibility and scattering processes.}

Now we consider the scattering of a particle by a localized single
scatterer. We assume for simplicity that the matrix elements $V_{\overline{p}%
\,\overline{p}^{\prime }}=\langle \overline{p}|\widehat{V}|\overline{p}%
^{\prime }\rangle $ of the potential in the basis of eigenvectors of the
momentum are well behaved ordinary functions of $\overline{p}$ and $%
\overline{p}^{\prime }$.

The Hamiltonian of the system is 
\begin{equation}
H=H_o+V=\int d\overline{p}\,\varepsilon _p|\overline{p}\rangle \langle 
\overline{p}|+\int \int d\overline{p}\,d\overline{p}^{\prime }\,V_{\overline{%
p}\,\overline{p}^{\prime }}|\overline{p}\rangle \langle \overline{p}^{\prime
}|,\qquad \varepsilon _p=\frac{p^2}{2m}.  \label{3.1}
\end{equation}

Let us consider observables $O\in {\cal O}$ of the form: 
\begin{equation}
O=\int d\overline{p}\,O_{\overline{p}}|\overline{p}\rangle \langle \overline{%
p}|+\int \int d\overline{p}\,d\overline{p}^{\prime }\,O_{\overline{p}\,%
\overline{p}^{\prime }}|\overline{p}\rangle \langle \overline{p}^{\prime }|,
\label{3.2}
\end{equation}
where $O_{\overline{p}}$ and $O_{\overline{p}\,\overline{p}^{\prime }}$ are
two independent regular functions of the variables $\overline{p}$ and $%
\overline{p}^{\prime }$, satisfying 
\[
O_{\overline{p}}^{*}=O_{\overline{p}},\qquad O_{\overline{p}\,\overline{p}%
^{\prime }}^{*}=O_{\overline{p}^{\prime }\,\overline{p}}. 
\]
Precisely the Hamiltonian (\ref{3.1}) is of the form given by (\ref{3.2})%
\footnote{%
This is a suitable choice for this problem in which we have a single
particle. In this case observables like momentum or energy have a diagonal
part as in equation (\ref{3.2}) and should have well defined mean values. It
is not the case in the thermodynamic limit, where extensive observables have
infinite mean values \cite{laura}.}.

As we stated in the previous section, the states are represented by
antilinear functionals acting on observables. In this case, a state $\rho $
will be represented by two regular functions $\rho _{\overline{p}}$ and $%
\rho _{\overline{p}\overline{p}^{\prime }}$. The mean value of an observable 
$O$ is 
\begin{equation}
(\rho |O)=\int d\overline{p}\,\rho _{\overline{p}}^{*}O_{\overline{p}}+\int
\int d\overline{p}\,d\overline{p}^{\prime }\,\rho _{\overline{p}\,\overline{p%
}^{\prime }}^{*}O_{\overline{p}\,\overline{p}^{\prime }}.  \label{3.3}
\end{equation}

From the conditions of total probability and reality of the mean values,
given by equations (\ref{2.7}) and (\ref{2.8}) we obtain: 
\begin{equation}
\rho _{\overline{p}}^{*}=\rho _{\overline{p}}\geq 0,\quad \rho _{\overline{p}%
\,\overline{p}^{\prime }}^{*}=\rho _{\overline{p}^{\prime }\,\overline{p}%
},\quad \int d\overline{p}\,\rho _{\overline{p}}^{*}=1  \label{3.4}
\end{equation}

It is useful to use a special notation for the generalized observables
expanding ${\cal O}$. Therefore we define: 
\begin{equation}
|\overline{p})\equiv |\overline{p}\rangle \langle \overline{p}|,\qquad
\qquad |\overline{p}\,\overline{p}^{\prime })\equiv |\overline{p}\rangle
\langle \overline{p}^{\prime }|,  \label{3.5}
\end{equation}

We also define the functionals $(\overline{p}|$ and $(\overline{p}\,%
\overline{p}^{\prime }|$ satisfying 
\begin{eqnarray}
(\overline{p}|\overline{k}) &=&\delta ^3(\overline{p}-\overline{k}), 
\nonumber \\
(\overline{p}\,\overline{p}^{\prime }|\overline{k}\,\overline{k}^{\prime })
&=&\delta ^3(\overline{p}-\overline{k})\delta ^3(\overline{p}^{\prime }-%
\overline{k}^{\prime }),  \label{3.6} \\
(\overline{p}|\overline{k}\,\overline{k}^{\prime }) &=&(\overline{p}\,%
\overline{p}^{\prime }|\overline{k})=0.  \nonumber
\end{eqnarray}

These 'basis' can be used to expand states and observables 
\begin{eqnarray}
\rho &=&\int d\overline{p}\,\rho _{\overline{p}}^{*}(\overline{p}|+\int \int
d\overline{p}\,d\overline{p}^{\prime }\,\rho _{\overline{p}\,\overline{p}%
^{\prime }}^{*}(\overline{p}\,\overline{p}^{\prime }|  \nonumber \\
O &=&\int d\overline{p}\,O_{\overline{p}}|\overline{p})+\int \int d\overline{%
p}\,d\overline{p}^{\prime }\,O_{\overline{p}\,\overline{p}^{\prime }}|%
\overline{p}\,\overline{p}^{\prime }).  \label{3.7}
\end{eqnarray}
With (\ref{3.6}) and (\ref{3.7}) we can deduce (\ref{3.3})\footnote{%
At this stage the formalism may look rather exotic. To make contact with
more usual things, let us mention that a pure state represented by the
normalized wave function $|\varphi \rangle =\int d\overline{k\,}\varphi _{%
\overline{k}}\,|\overline{k}\rangle $, can also be represented by the
functional $\rho =\int d\overline{p}\varphi _{\overline{p}}^{*}\varphi _{%
\overline{p}}(\overline{p}|+\int \int d\overline{p}\,d\overline{p}^{\prime
}\varphi _{\overline{p}}^{*}\varphi _{\overline{p}^{\prime }}(\overline{p}\,%
\overline{p}^{\prime }|$}.

Starting from the definitions 
\[
{\Bbb L}_0^{\dagger }O=[H_0,O],\qquad {\Bbb L}_V^{\dagger }O=[V,O], 
\]
and using the 'basis' $|\overline{p})$, $|\overline{p}\,\overline{p}^{\prime
})$ for the observables and $(\overline{p}|$, $(\overline{p}\,\overline{p}%
^{\prime }|$ for the states, we obtain: 
\begin{eqnarray}
{\Bbb L}_0^{\dagger } &=&\int d\overline{p}\,d\overline{p}^{\prime
}\,(\varepsilon _{\overline{p}}-\varepsilon _{\overline{p}^{\prime }})\,|%
\overline{p}\,\overline{p}^{\prime })(\overline{p}\,\overline{p}^{\prime }|,
\nonumber \\
{\Bbb L}_V^{\dagger } &=&\int d\overline{p}\,d\overline{p}^{\prime }\,|%
\overline{p}\,\overline{p}^{\prime })V_{\overline{p}\,\overline{p}^{\prime
}}\,\left[ (\overline{p}^{\prime }|-(\overline{p}|\right]  \nonumber \\
&&+\int d\overline{p}\,d\overline{p}^{\prime }\,|\overline{p}\,\overline{p}%
^{\prime })\,\int d\overline{p}^{\prime \prime }\left[ V_{\overline{p}\,%
\overline{p}^{\prime \prime }}(\overline{p}^{\prime \prime }\,\overline{p}%
^{\prime }|-V_{\overline{p}^{\prime \prime }\,\overline{p}^{\prime }}(%
\overline{p}\,\overline{p}^{\prime \prime }|\right]  \label{3.8}
\end{eqnarray}

For the projectors ${\Bbb P}^{\dagger }$ and ${\Bbb Q}^{\dagger }$, used in
the previous section to decompose the resolvent, we choose in this section
the following explicit form: 
\[
{\Bbb P}^{\dagger }O=\int d\overline{p}\,O_{\overline{p}}|\overline{p}%
\rangle \langle \overline{p}|,\qquad {\Bbb Q}^{\dagger }O=\int \int d%
\overline{p}\,d\overline{p}^{\prime }\,O_{\overline{p}\,\overline{p}^{\prime
}}|\overline{p}\rangle \langle \overline{p}^{\prime }|, 
\]
or equivalently 
\begin{equation}
{\Bbb P}^{\dagger }=\int d\overline{p}\,|\overline{p})(\overline{p}|,\qquad 
{\Bbb Q}^{\dagger }=\int d\overline{p}\,d\overline{p}^{\prime }\,|\overline{p%
}\,\overline{p}^{\prime })(\overline{p}\,\overline{p}^{\prime }|.
\label{3.9}
\end{equation}

The superoperators ${\Bbb P}^{\dagger }$ and ${\Bbb Q}^{\dagger }$ project
the observables into their diagonal and off diagonal parts. The
corresponding superoperators ${\Bbb P}$ and ${\Bbb Q}$ project the states
into their diagonal and off diagonal parts, i.e. 
\begin{equation}
({\Bbb P}\rho |=\int d\overline{p}\,\rho _{\overline{p}}^{*}(\overline{p}%
|,\qquad ({\Bbb Q}\rho |=\int \int d\overline{p}\,d\overline{p}^{\prime
}\,\rho _{\overline{p}\,\overline{p}^{\prime }}^{*}(\overline{p}\,\overline{p%
}^{\prime }|.  \label{3.10}
\end{equation}

From (\ref{3.8}) and (\ref{3.9}) we obtain 
\begin{equation}
{\Bbb P}^{\dagger }{\Bbb L}^{\dagger }=0,\qquad \Psi ^{\dagger }(z)=0,\qquad 
{\Bbb C}^{\dagger }(z)=0,  \label{3.10b}
\end{equation}
and the decomposition of the resolvent reduces to 
\begin{equation}
\frac 1{{\Bbb L}^{\dagger }+z}=\frac 1z{\Bbb P}^{\dagger }{\Bbb +}\frac 1z%
{\Bbb Q}^{\dagger }{\Bbb D}^{\dagger }(z){\Bbb P}^{\dagger }+{\Bbb Q}%
^{\dagger }\frac 1{{\Bbb Q}^{\dagger }{\Bbb L}^{\dagger }{\Bbb Q}^{\dagger }%
{\Bbb +}z}{\Bbb Q}^{\dagger }.  \label{3.11}
\end{equation}

The time evolution is given by 
\begin{eqnarray}
({\Bbb P}\rho _t| &=&\frac i{2\pi }\int\limits_\Gamma dz\,\frac{\exp (-izt)}z%
({\Bbb P}\rho _o|  \nonumber \\
&&-\frac i{2\pi }\int\limits_\Gamma dz\,\frac{\exp (-izt)}z({\Bbb Q}\rho _o|%
\frac 1{{\Bbb Q}^{\dagger }{\Bbb L}^{\dagger }{\Bbb Q}^{\dagger }{\Bbb +}z}%
{\Bbb Q}^{\dagger }{\Bbb L}^{\dagger }{\Bbb P}^{\dagger }  \label{3.12} \\
({\Bbb Q}\rho _t| &=&\frac i{2\pi }\int\limits_\Gamma dz\,\exp (-izt)({\Bbb Q%
}\rho _o|\frac 1{{\Bbb Q}^{\dagger }{\Bbb L}^{\dagger }{\Bbb Q}^{\dagger }%
{\Bbb +}z}{\Bbb Q}^{\dagger }  \label{3.13}
\end{eqnarray}

Equation (\ref{3.12}) shows the influence of the diagonal part $({\Bbb P}%
\rho _o|$ and the off diagonal part $({\Bbb Q}\rho _o|$ of the initial state
on the diagonal part of the state at time $t$. As the collision operator $%
\Psi ^{\dagger }(z)$ is zero, there are no diagonal-diagonal transitions in
the process.

Equation (\ref{3.13}) shows that there is no influence of the diagonal part
of the initial condition on the off diagonal part of the state at time $t$.

It is easy to show that the first factor in (\ref{3.12}) is time
independent. The integral over the horizontal line $\Gamma $ in the upper
half plane can be closed over a very big semicircle in the lower half plane.
The integral over this big semicircle has vanishing contribution when the
radius goes to infinity. Then the closed curve can be deformed into a small
circle around the sinple pole at $z=0$, to obtain 
\begin{equation}
\frac i{2\pi }\int\limits_\Gamma dz\,\frac{\exp (-izt)}z({\Bbb P}\rho _o|O)=(%
{\Bbb P}\rho _o|{\Bbb P}^{\dagger }O),  \label{3.14}
\end{equation}
for all observables $O$.

To analyze the second factor it is convenient to use the complete
biorthogonal system of Lipmann-Schwinger generalized eigenvectors of the
Hamiltonian 
\begin{eqnarray}
|\overline{k}^{\pm }\rangle &=&|\overline{k}\rangle +\frac 1{\varepsilon
_k\pm i0-H}V|\overline{k}\rangle ,  \nonumber \\
\langle \overline{k}^{\pm }| &=&\langle \overline{k}|+\langle \overline{k}|V%
\frac 1{\varepsilon _k\mp i0-H},  \label{3.15}
\end{eqnarray}
for which 
\[
I=\int d\overline{k}\,|\overline{k}^{\pm }\rangle \langle \overline{k}^{\pm
}|,\qquad H=\int d\overline{k}\,\varepsilon _k\,|\overline{k}^{\pm }\rangle
\langle \overline{k}^{\pm }|,\qquad \langle \overline{k}^{\pm }|\overline{k}%
^{\prime \pm }\rangle =\delta ^3(\overline{k}-\overline{k}^{\prime }). 
\]

We have 
\begin{eqnarray}
&&-\frac i{2\pi }\int\limits_\Gamma dz\,\frac{\exp (-izt)}z({\Bbb Q}\rho _o|%
\frac 1{{\Bbb Q}^{\dagger }{\Bbb L}^{\dagger }{\Bbb Q}^{\dagger }{\Bbb +}z}%
{\Bbb Q}^{\dagger }{\Bbb L}^{\dagger }{\Bbb P}^{\dagger }|O)  \nonumber \\
&=&-\frac i{2\pi }\int\limits_\Gamma dz\frac{\exp (-izt)}z\int \int d%
\overline{k}d\overline{k}^{\prime }  \nonumber \\
&&({\Bbb Q}\rho _o|\frac 1{{\Bbb Q}^{\dagger }{\Bbb L}^{\dagger }{\Bbb Q}%
^{\dagger }{\Bbb +}z}\left[ |\overline{k}^{+}\rangle \langle \overline{k}%
^{+}|[H,{\Bbb P}^{\dagger }O]|\overline{k}^{\prime +}\rangle \langle 
\overline{k}^{\prime +}|\right] )  \nonumber \\
&=&-\frac i{2\pi }\int\limits_\Gamma dz\frac{\exp (-izt)}z\int \int d%
\overline{k}d\overline{k}^{\prime }  \nonumber \\
&&\frac{\varepsilon _k-\varepsilon _{k^{\prime }}}{\varepsilon
_k-\varepsilon _{k^{\prime }}+z}({\Bbb Q}\rho _o||\overline{k}^{+}\rangle
\langle \overline{k}^{\prime +}|)\langle \overline{k}^{+}|{\Bbb P}^{\dagger
}O|\overline{k}^{\prime +}\rangle .  \label{3.16}
\end{eqnarray}
The integral over $\overline{k}$ and $\overline{k}^{\prime }$ can be
transformed using polar coordinates into an integral over $\varepsilon _k$, $%
\varepsilon _{k^{\prime }}$ and over the angles. Looking at the integrals
over the energies we can write 
\begin{eqnarray*}
&&\int\limits_\Gamma dz\,\frac{\exp (-izt)}z\int\limits_0^\infty
d\varepsilon \int\limits_0^\infty d\varepsilon ^{\prime }\frac{\varepsilon
-\varepsilon ^{\prime }}{\varepsilon -\varepsilon ^{\prime }+z}%
\,f(\varepsilon _k,\varepsilon _{k^{\prime }}) \\
&=&-\int\limits_\Gamma dz\,\frac{\exp (-izt)}z\int\limits_0^\infty d\lambda
\int\limits_{-\lambda }^\lambda d\nu \frac \nu {z-\nu }\,f(\lambda ,\nu ) \\
&=&-\int\limits_0^\infty d\lambda \int\limits_\Gamma dz\,\frac{\exp (-izt)}z%
\,F_\lambda (z).
\end{eqnarray*}
where for the last expressions we used the variables $\nu =\varepsilon
^{\prime }-\varepsilon $ and $\lambda =\frac 12(\varepsilon ^{\prime
}+\varepsilon )$, and $F_\lambda (z)\equiv \int_{-\lambda }^\lambda d\nu 
\frac \nu {z-\nu }\,f(\lambda ,\nu )$. In terms of the complex variable $z$, 
$\frac{\exp (-izt)}z\,F_\lambda (z)$ has a simple pole in $z=0$ and a cut in
the real interval $(-\lambda ,+\lambda )$: 
\[
F_\lambda (x+i0)-F_\lambda (x-i0)= 
\begin{array}{l}
0\quad \quad \quad \quad \quad \quad \;if\;x\notin (-\lambda ,+\lambda ) \\ 
-2\pi i\,x\,f(\lambda ,x)\quad \,if\;x\in (-\lambda ,+\lambda )
\end{array}
\]

As in the previous case, the integral over $\Gamma $ can be closed in the
lower half plane, surrounding the pole and the cut. Using a closed curve
very close to the cut we obtain 
\begin{eqnarray}
&&-\int\limits_0^\infty d\lambda \int\limits_\Gamma dz\,\frac{\exp (-izt)}z%
\,F_\lambda (z)  \nonumber \\
&=&2\pi i\left\{ \int\limits_0^\infty d\lambda \int\limits_0^\infty d\nu
\,f(\lambda ,\nu )+\int\limits_0^\infty d\lambda \int\limits_0^\infty d\nu
\exp (-i\nu t)\,\,\,f(\lambda ,\nu )\right\}  \label{3.17}
\end{eqnarray}

We can now insert (\ref{3.17}) in (\ref{3.16}) to obtain 
\begin{eqnarray*}
&&-\frac i{2\pi }\int\limits_\Gamma dz\,\frac{\exp (-izt)}z({\Bbb Q}\rho _o|%
\frac 1{{\Bbb Q}^{\dagger }{\Bbb L}^{\dagger }{\Bbb Q}^{\dagger }{\Bbb +}z}%
{\Bbb Q}^{\dagger }{\Bbb L}^{\dagger }{\Bbb P}^{\dagger }|O) \\
&=&\int \int d\overline{k}\,d\overline{k}^{\prime }\,\exp \{i(\varepsilon
_k-\varepsilon _{k^{\prime }})t\}\,\,\,({\Bbb Q}\rho _o||\overline{k}%
^{+}\rangle \langle \overline{k}^{+}|{\Bbb P}^{\dagger }O|\overline{k}%
^{\prime +}\rangle \langle \overline{k}^{\prime +}|).
\end{eqnarray*}

Therefore 
\begin{eqnarray}
({\Bbb P}\rho _t|O) &=&({\Bbb P}\rho _o|{\Bbb P}^{\dagger }O)+  \nonumber \\
&&\int \int d\overline{k}d\overline{k}^{\prime }\exp \{i(\varepsilon
_k-\varepsilon _{k^{\prime }})t\}({\Bbb Q}\rho _o||\overline{k}^{+}\rangle
\langle \overline{k}^{+}|{\Bbb P}^{\dagger }O|\overline{k}^{\prime +}\rangle
\langle \overline{k}^{\prime +}|)  \label{3.18}
\end{eqnarray}

The Lipmann-Schwinger vectors (\ref{3.15}) can also be used in expression (%
\ref{3.13}) 
\begin{equation}
({\Bbb Q}\rho _t|O)=\int \int d\overline{k}d\overline{k}^{\prime }\exp
\{i(\varepsilon _k-\varepsilon _{k^{\prime }})t\}({\Bbb Q}\rho _o||\overline{%
k}^{+}\rangle \langle \overline{k}^{+}|{\Bbb Q}^{\dagger }O|\overline{k}%
^{\prime +}\rangle \langle \overline{k}^{\prime +}|)  \label{3.19}
\end{equation}

There are no singular terms in $({\Bbb Q}\rho _o||\overline{k}^{+}\rangle
\langle \overline{k}^{+}|{\Bbb Q}^{\dagger }O|\overline{k}^{\prime +}\rangle
\langle \overline{k}^{\prime +}|)$, and therefore Riemann- Lebesgue theorem
can be used in (\ref{3.19}) to obtain 
\begin{equation}
\lim_{t\rightarrow \infty }({\Bbb Q}\rho _t|O)=0.  \label{3.20}
\end{equation}

From equations (\ref{3.18}) and (\ref{3.19}) it is easy to show that there
is no time evolution for a diagonal initial condition, i.e. 
\[
\rho _o={\Bbb P}\rho _o\Longrightarrow \rho _t=\rho _o. 
\]

We must isolate the singular term in $({\Bbb Q}\rho _o||\overline{k}%
^{+}\rangle \langle \overline{k}^{+}|{\Bbb P}^{\dagger }O|\overline{k}%
^{\prime +}\rangle \langle \overline{k}^{\prime +}|)$ before using the
Riemann-Lebesgue theorem to compute $\lim_{t\rightarrow \infty }({\Bbb P}%
\rho _t|O)$. Using (\ref{3.15}) we obtain: 
\begin{eqnarray}
&&({\Bbb Q}\rho _o||\overline{k}^{+}\rangle \langle \overline{k}^{+}|{\Bbb P}%
^{\dagger }O|\overline{k}^{\prime +}\rangle \langle \overline{k}^{\prime +}|)
\nonumber \\
&=&({\Bbb Q}\rho _o||\overline{k}^{+}\rangle \langle \overline{k}^{\prime
+}|)\times \{O_{\overline{k}}\delta ^3(\overline{k}-\overline{k}^{\prime
})+O_{\overline{k}}\langle \overline{k}|\frac 1{\varepsilon _{k^{\prime
}}+i0-H}V|\overline{k}^{\prime }\rangle +  \nonumber \\
&&+\langle \overline{k}|V\frac 1{\varepsilon _{k^{\prime }}-i0-H}|\overline{k%
}^{\prime }\rangle O_{\overline{k}^{\prime }}+  \nonumber \\
&&+\langle \overline{k}|V\frac 1{\varepsilon _{k^{\prime }}-i0-H}|\overline{k%
}^{\prime \prime }\rangle O_{\overline{k}^{\prime \prime }}\langle \overline{%
k}^{\prime \prime }|\frac 1{\varepsilon _{k^{\prime }}+i0-H}V|\overline{k}%
^{\prime }\rangle \}.  \label{3.21}
\end{eqnarray}

Replacing (\ref{3.21}) in (\ref{3.18}), and using the Riemann-Lebesgue
theorem we obtain: 
\begin{eqnarray}
\lim_{t\rightarrow \infty }({\Bbb P}\rho _t|O) &=&({\Bbb P}\rho _o|{\Bbb P}%
^{\dagger }O)+\int d\overline{k}\,({\Bbb Q}\rho _o||\overline{k}^{+}\rangle
\langle \overline{k}^{+}|)O_{\overline{k}}  \nonumber \\
&=&\int d\overline{k}\,(\rho _o||\overline{k}^{+}\rangle \langle \overline{k}%
^{+}|)O_{\overline{k}}.  \label{3.22}
\end{eqnarray}

Therefore, in weak sense: 
\begin{equation}
(\rho _\infty |=\lim_{t\rightarrow \infty }(\rho _t|=\int d\overline{k}%
\,(\rho _o||\overline{k}^{+}\rangle \langle \overline{k}^{+}|)(\overline{k}|.
\label{3.23}
\end{equation}

This result shows a sort of ''weak intrinsic irreversibility'' of the
scattering process. As we mentioned, a pure state which can be represented
by a normalizable wave function $|\varphi \rangle =\int d\overline{k}%
\,\varphi _{\overline{k}}\,|\overline{k}\rangle $, can also be represented
by the functional $\rho =\int d\overline{p}\varphi _{\overline{p}%
}^{*}\varphi _{\overline{p}}(\overline{p}|+\int \int d\overline{p}\,d%
\overline{p}^{\prime }\varphi _{\overline{p}}^{*}\varphi _{\overline{p}%
^{\prime }}(\overline{p}\,\overline{p}^{\prime }|$. Therefore, in this
formalism, $\rho _{\overline{k}}^{*}=\rho _{\overline{k}\overline{k}^{\prime
}}^{*}$ is a necessary condition to have a pure state. We are used to the
idea that the 'purity' of a state is preserved by the time evolution.
However, equation (\ref{3.23}) states that, in weak sense and for all
initial conditions, the evolution is not towards a pure state but towards a
'generalized mixture' in which $(\rho _\infty )_{\overline{k}}^{*}=(\rho _o||%
\overline{k}^{+}\rangle \langle \overline{k}^{+}|)$ and $(\rho _\infty )_{%
\overline{k}\overline{k}^{\prime }}^{*}=0$. This 'final' state is time
invariant, because from (\ref{3.10b}) we have 
\[
({\Bbb L}\rho _\infty |O)=({\Bbb LP}\rho _\infty |O)=(\rho _\infty |{\Bbb P}%
^{\dagger }{\Bbb L}^{\dagger }O)=0\;\Longrightarrow \;{\Bbb L}\rho _\infty
=0. 
\]
Moreover, the time inversion ${\Bbb T}\rho _\infty $ of the 'final' state is
also invariant under time evolution. In fact, for any $\rho =\int d\overline{%
p}\,\rho _{\overline{p}}^{*}(\overline{p}|+\int \int d\overline{p}\,d%
\overline{p}^{\prime }\,\rho _{\overline{p}\,\overline{p}^{\prime }}^{*}(%
\overline{p}\,\overline{p}^{\prime }|$ we have 
\[
{\Bbb T}\rho =\int d\overline{p}\,\rho _{-\overline{p}}(\overline{p}|+\int
\int d\overline{p}\,d\overline{p}^{\prime }\,\rho _{-\overline{p}\,-\,%
\overline{p}^{\prime }}(\overline{p}\,\overline{p}^{\prime }|, 
\]
then 
\[
{\Bbb T}\rho _\infty =\int d\overline{k}\,(\rho _o||-\overline{k}^{+}\rangle
\langle -\overline{k}^{+}|)^{*}(\overline{k}|, 
\]
and 
\[
{\Bbb LT}\rho _\infty =0. 
\]

Therefore, the time evolution of the time inversion of the 'final' state
cannot reproduce the initial state. We may say, in this sense, that the
scattering process is intrinsically irreversible. But this irreversibility
appears for processes involving an infinite amount of time, as the 'final'
state is obtained with $t\rightarrow \infty $. For a very big time $%
t_o<\infty $, the time inversion is possible in principle, although it may
be very difficult to prepare the state ${\Bbb T}\rho _{t_o}$ in practice.

\section{Real and complex spectral decompositions.}

In the previous section we used the formalism of states and observables with
diagonal singularities to obtain the time evolution of generalized states
(equations (\ref{3.18}) and (\ref{3.19})). A real spectral decomposition of
the Liouville-Von Neumann superoperator is implicit in these equations.

This real spectral decomposition was enough to compute the 'final' state and
to argue on the intrinsic irreversibility of the process. In this section we
are going to make explicit trough the spectral decomposition the influence
of the resonances produced by the poles of the analytic extensions of the
resolvent, which in scattering process are determined by the poles of the '$%
S $ matrix'.

Although it is possible to compute the complex spectral decomposition for
the model of the previous section, we prefer to analyze a simplified model.

Let us consider a system with Hamiltonian 
\begin{equation}
H=H_0+V=\int\limits_0^\infty d\omega \,\omega \,|\omega \rangle \langle
\omega |+\int\limits_0^\infty d\omega \int\limits_0^\infty d\omega ^{\prime
}V_{\omega \omega ^{\prime }}\,|\omega \rangle \langle \omega ^{\prime }|,
\label{4.1}
\end{equation}
where the generalized right (left) eigenvectors $|\omega \rangle $ ($\langle
\omega |$)of $H_0$ form a complete biorthonormal system 
\[
I=\int\limits_0^\infty d\omega \,|\omega \rangle \langle \omega |,\qquad
\langle \omega |\omega ^{\prime }\rangle =\delta (\omega -\omega ^{\prime
}). 
\]

We also assume that the Lipmann-Schwinger generalized eigenvectors of the
Hamiltonian 
\begin{eqnarray}
|\omega ^{\pm }\rangle &=&|\omega \rangle +\frac 1{\omega \pm i0-H}V|\omega
\rangle  \nonumber \\
\langle \omega ^{\pm }| &=&\langle \omega |+\langle \omega |V\frac 1{\omega
\mp i0-H}  \label{4.2}
\end{eqnarray}
are also a complete set: 
\[
H=\int\limits_0^\infty d\omega \,\omega \,|\omega ^{\pm }\rangle \langle
\omega ^{\pm }|,\qquad I=\int\limits_0^\infty d\omega \,|\omega ^{\pm
}\rangle \langle \omega ^{\pm }|,\qquad \langle \omega ^{\pm }|\omega
^{\prime \pm }\rangle =\delta (\omega -\omega ^{\prime }). 
\]

The vectors $|\omega ^{+}\rangle $ and $|\omega ^{-}\rangle $ are related by
the ''$S$ matrix'' 
\begin{equation}
|\omega ^{+}\rangle =S(\omega )|\omega ^{-}\rangle ,\qquad \langle \omega
^{+}|=S^{*}(\omega )\langle \omega ^{-}|,  \label{4.3}
\end{equation}
where 
\begin{eqnarray}
S(\omega ) &=&1-2\pi i\langle \omega |V|\omega \rangle -2\pi i\langle \omega
|V\frac 1{\omega +i0-H}V|\omega \rangle ,  \nonumber \\
S^{*}(\omega ) &=&1+2\pi i\langle \omega |V|\omega \rangle +2\pi i\langle
\omega |V\frac 1{\omega -i0-H}V|\omega \rangle .  \label{4.4}
\end{eqnarray}

We also assume that the analytic extension $\left( \frac 1{s-H}\right)
_{s=z}^{\pm }$ of the resolvent $\frac 1{z-H}$, from the upper (lower) to
the lower (upper) half plane, has a simple pole at $z=z_0$ ($z=z_0^{*}$),
where $%
\mathop{\rm Im}
z_0<0$ ($%
\mathop{\rm Im}
z_0^{*}>0$).

From (\ref{4.4}) we can define the following analytic extensions 
\begin{eqnarray}
S(z) &=&1-2\pi i\langle z|V|z\rangle -2\pi i\langle z|V\left( \frac 1{s-H}%
\right) _{s=z}^{+}V|z\rangle ,  \nonumber \\
S^{*}(z) &=&1+2\pi i\langle z|V|z\rangle +2\pi i\langle z|V\left( \frac 1{s-H%
}\right) _{s=z}^{-}V|z\rangle .  \label{4.5}
\end{eqnarray}

The analytic extension $S(z)$ ($S^{*}(z)$) of $S(\omega )$ ($S^{*}(\omega )$%
) has a simple pole at $z=z_0$ ($z=z_0^{*}$).

From the Lipmann-Schwinger generalized eigenvectors of the Hamiltonian we
may also construct the corresponding analytic extensions 
\begin{eqnarray}
|z^{\pm }\rangle &=&|z\rangle +\left( \frac 1{s-H}\right) _{s=z}^{\pm
}V|z\rangle  \nonumber \\
\langle z^{\pm }| &=&\langle z|+\langle z|V\left( \frac 1{s-H}\right)
_{s=z}^{\mp }.  \label{4.6}
\end{eqnarray}
In the previous expresions, $|z\rangle $ and $\langle z|$ are functionals
defined by 
\[
\langle z|\varphi \rangle =\varphi (z),\qquad \langle \varphi |z\rangle
=\varphi ^{*}(z), 
\]
where $\varphi (z)$ and $\varphi ^{*}(z)$ are the analytic extensions of $%
\varphi (\omega )=\langle \omega |\varphi \rangle $ and $\varphi ^{*}(\omega
)=\langle \varphi |\omega \rangle $.

As in section 3,we consider observables of the form 
\begin{equation}
O={\Bbb P}^{\dagger }O+{\Bbb Q}^{\dagger }O=\int d\omega O_\omega |\omega
)+\int \int d\omega d\omega ^{\prime }O_{\omega \omega ^{\prime }}|\omega
\omega ^{\prime }),  \label{4.7}
\end{equation}
\[
|\omega )\equiv |\omega \rangle \langle \omega |,\;|\omega \omega ^{\prime
})\equiv |\omega \rangle \langle \omega ^{\prime }|,\quad O_\omega =O_\omega
^{*},\quad O_{\omega \omega ^{\prime }}=O_{\omega ^{\prime }\omega }^{*}. 
\]

If we define the functionals $(\omega |$ and $(\omega \omega ^{\prime }|$ by
the following equations 
\begin{eqnarray*}
(\omega |\omega ^{\prime }) &=&\delta (\omega -\omega ^{\prime }) \\
(\omega |\omega ^{\prime }\omega ^{\prime \prime }) &=&(\omega ^{\prime
}\omega ^{\prime \prime }|\omega )=0 \\
(\omega \omega ^{\prime }|\varsigma \varsigma ^{\prime }) &=&\delta (\omega
-\varsigma )\delta (\omega ^{\prime }-\varsigma ^{\prime }),
\end{eqnarray*}
they can be used to represent the state functionals 
\begin{eqnarray*}
\rho &=&\int d\omega \rho _\omega ^{*}(\omega |+\int \int d\omega d\omega
^{\prime }\rho _{\omega \omega ^{\prime }}^{*}(\omega \omega ^{\prime }| \\
\rho _\omega ^{*} &=&\rho _\omega \geq 0,\quad \rho _{\omega \omega ^{\prime
}}^{*}=\rho _{\omega ^{\prime }\omega },\quad \int d\omega \rho _\omega
^{*}=1.
\end{eqnarray*}

The mean value of an observable $O$ in the state $\rho $ is given by 
\[
\langle O\rangle _\rho =(\rho |O)=\int d\omega \rho _\omega ^{*}O_\omega
+\int \int d\omega d\omega ^{\prime }\rho _{\omega \omega ^{\prime
}}^{*}O_{\omega \omega ^{\prime }}, 
\]
with the following time evolution 
\[
(\rho _t|O)=(U_t\rho _0|O)=(\rho _0|U_t^{\dagger }O)=(\rho _0|\exp
(iHt)O\exp (-iHt)). 
\]

The observable given in (\ref{4.7}) can be written as 
\begin{eqnarray*}
O &=&O_{diag}+O_{reg} \\
O_{diag} &=&\int d\omega O_\omega |\omega \rangle \langle \omega | \\
O_{reg} &=&\int \int d\omega d\omega ^{\prime }O_{\omega \omega ^{\prime
}}|\omega \rangle \langle \omega ^{\prime }|
\end{eqnarray*}

Using the Lipmann-Schwinger vectors (\ref{4.2}) we have 
\[
O_{diag}=O_{inv}+\Delta O_{diag}, 
\]
where 
\begin{eqnarray*}
O_{inv} &=&\int d\omega O_\omega |\omega ^{+}\rangle \langle \omega ^{+}| \\
\Delta O_{diag} &=&\int d\omega O_\omega (|\Delta \omega \rangle \langle
\Delta \omega |+|\Delta \omega \rangle \langle \omega ^{+}|+|\omega
^{+}\rangle \langle \Delta \omega |) \\
|\Delta \omega \rangle &=&|\omega \rangle -|\omega ^{+}\rangle =-\frac 1{%
\omega +i0-H}V|\omega \rangle .
\end{eqnarray*}

Therefore the observable $O$ can be decomposed into an invariant ($inv$) and
a fluctuating ($fluc$) part 
\begin{eqnarray}
O &=&O_{inv}+O_{fluc}  \nonumber \\
O_{fluc} &=&O_{reg}+\Delta O_{diag}.  \label{4.8}
\end{eqnarray}
The fluctuating part $O_{fluc}$ has no diagonal singularity ($%
(O_{fluc})_\omega =0$), and $O_{inv}$ is time independent ($U_t^{\dagger
}O_{inv}=O_{inv}$).

The real spectral decomposition of the time evolution can be obtained using
the decomposition $O=O_{inv}+O_{fluc}$ of the observables and the
Lipmann-Schwinger vectors (\ref{4.2}): 
\begin{eqnarray}
(\rho _t|O) &=&(\rho _0|O_{inv})+(\rho _0|\exp [iHt]O_{fluc}\exp [-iHt]) 
\nonumber \\
&=&\int_0^\infty d\omega (\rho _0||\omega ^{+}\rangle \langle \omega
^{+}|)O_\omega +  \nonumber \\
&&+\int_0^\infty d\omega \int_0^\infty d\omega ^{\prime }e^{i(\omega -\omega
^{\prime })t}(\rho _0||\omega ^{+}\rangle \langle \omega
^{+}|O_{fluc}|\omega ^{\prime +}\rangle \langle \omega ^{\prime +}|)
\label{4.9}
\end{eqnarray}

The last term will vanish when $t\rightarrow \infty $ and therefore, in weak
sense 
\begin{equation}
(\rho _\infty |=\lim\limits_{t\rightarrow \infty }(\rho _t|=\int_0^\infty
d\omega (\rho _0||\omega ^{+}\rangle \langle \omega ^{+}|)(\omega |.
\label{4.10}
\end{equation}

Expression (\ref{4.9}) correspond to the following real spectral
decomposition of the identity (${\Bbb I}^{\dagger }$) and Liouville-Von
Neumann (${\Bbb L}^{\dagger }$) superoperators 
\begin{eqnarray}
{\Bbb I}^{\dagger } &=&\int_0^\infty d\omega |\Phi _\omega )(\widetilde{\Phi 
}_\omega |+\int_0^\infty d\omega \int_0^\infty d\omega ^{\prime }|\Phi
_{\omega \omega ^{\prime }})(\widetilde{\Phi }_{\omega \omega ^{\prime }}|, 
\nonumber \\
{\Bbb L}^{\dagger } &=&\int_0^\infty d\omega \int_0^\infty d\omega ^{\prime
}\,(\omega -\omega ^{\prime })|\Phi _{\omega \omega ^{\prime }})(\widetilde{%
\Phi }_{\omega \omega ^{\prime }}|  \label{4.12}
\end{eqnarray}
where 
\begin{eqnarray}
|\Phi _\omega ) &=&||\omega ^{+}\rangle \langle \omega ^{+}|)  \nonumber \\
(\widetilde{\Phi }_\omega | &=&(\omega |  \nonumber \\
|\Phi _{\omega \omega ^{\prime }}) &=&||\omega ^{+}\rangle \langle \omega
^{\prime +}|)  \label{4.13} \\
(\widetilde{\Phi }_{\omega \omega ^{\prime }}| &=&\int dy\{\langle \omega
^{+}|y\rangle \langle y|\omega ^{\prime +}\rangle -\delta (\omega -y)\delta
(y-\omega ^{\prime })\}(y|+  \nonumber \\
&&+\int \int dydy^{\prime }\langle \omega ^{+}|y\rangle \langle y^{\prime
}|\omega ^{\prime +}\rangle (yy^{\prime }|  \nonumber
\end{eqnarray}

It is easy to prove that the generalized right and left eigenvectors of $%
{\Bbb L}^{\dagger }$ given in (\ref{4.13}) satisfy the orthogonality
conditions 
\begin{eqnarray*}
(\widetilde{\Phi }_\omega |\Phi _{\omega ^{\prime }}) &=&\delta (\omega
-\omega ^{\prime }) \\
(\widetilde{\Phi }_\omega |\Phi _{yy^{\prime }}) &=&(\widetilde{\Phi }%
_{\omega \omega ^{\prime }}|\Phi _y)=0 \\
(\widetilde{\Phi }_{\omega \omega ^{\prime }}|\Phi _{yy^{\prime }})
&=&\delta (\omega -y)\delta (\omega ^{\prime }-y^{\prime }).
\end{eqnarray*}

Any state can be expanded in terms of the complete biorthonormal system (\ref
{4.13}) 
\[
\rho =\int_0^\infty d\omega (\rho |\Phi _\omega )(\widetilde{\Phi }_\omega
|+\int_0^\infty d\omega \int_0^\infty d\omega ^{\prime }(\rho |\Phi _{\omega
\omega ^{\prime }})(\widetilde{\Phi }_{\omega \omega ^{\prime }}|, 
\]
and therefore it is important to give a physical meaning to the generalized
states $(\widetilde{\Phi }_\omega |$ and $(\widetilde{\Phi }_{\omega \omega
^{\prime }}|$. For the mean value of the total energy we obtain 
\[
(\widetilde{\Phi }_\omega |H)=(\omega |H)=\omega 
\]
\[
(\widetilde{\Phi }_{\omega \omega ^{\prime }}|H)= 
\]
\[
=\int dy\{\langle \omega ^{+}|y\rangle \langle y|\omega ^{\prime +}\rangle
-\delta (\omega -y)\delta (y-\omega ^{\prime })\}y+\int \int dydy^{\prime
}\langle \omega ^{+}|y\rangle V_{yy^{\prime }}\langle y^{\prime }|\omega
^{\prime +}\rangle =0, 
\]
and for the ''trace'' 
\[
(\widetilde{\Phi }_\omega |I)=(\omega |I)=(\omega |\int d\omega ^{\prime
}|\omega ^{\prime })=\int d\omega ^{\prime }\delta (\omega -\omega ^{\prime
})=1 
\]
\[
(\widetilde{\Phi }_{\omega \omega ^{\prime }}|I)= 
\]
\[
=\int dy\{\langle \omega ^{+}|y\rangle \langle y|\omega ^{\prime +}\rangle
-\delta (\omega -y)\delta (y-\omega ^{\prime })\}=\langle \omega ^{+}|\omega
^{\prime +}\rangle -\delta (\omega -\omega ^{\prime })=0. 
\]
The generalized state $(\widetilde{\Phi }_\omega |$ is a physical state with
energy $\omega $ and ''trace'' $1$, and $(\widetilde{\Phi }_{\omega \omega
^{\prime }}|$ has zero energy and zero ''trace''. Clearly, it is impossible
to realize a physical state including only $(\widetilde{\Phi }_{\omega
\omega ^{\prime }}|$ components.

The complex number $z_0^{*}$, where the analytic extension of $S^{*}(\omega
) $ has a simple pole, can be introduced in the spectral decomposition if,
in equation (\ref{4.9}), we deform the $\omega $ integral over ${\Bbb R}^{+}$
into an integral over a curve in the upper half plane, i.e. 
\begin{eqnarray*}
&&\int_0^\infty d\omega e^{i\omega t}|\omega ^{+}\rangle \langle \omega ^{+}|
\\
&=&\int_0^\infty d\omega e^{i\omega t}|\omega ^{+}\rangle \langle \omega
^{-}|S^{*}(\omega )\rightarrow \\
&\rightarrow &e^{iz_0^{*}t}|\widetilde{f_0}\rangle \langle
f_0|+\int_0^{-\infty }d\omega e^{i\omega t}|(\omega +i0)^{+}\rangle \langle
(\omega +i0)^{+}|
\end{eqnarray*}
\begin{equation}
|\widetilde{f_0}\rangle =[2\pi i(ResS^{*})_{z_0^{*}}]^{\frac 12%
}|z_0^{*+}\rangle ,\quad \langle f_0|=[2\pi i(ResS^{*})_{z_0^{*}}]^{\frac 12%
}\langle z_0^{*-}|.  \label{4.14}
\end{equation}

The complex numbers $z_0$, where the analytic extension of $S(\omega
^{\prime })$ has a simple pole, can be introduced in the spectral
decomposition if, in equation (\ref{4.9}), we deform the $\omega ^{\prime }$
integral over ${\Bbb R}^{+}$ into an integral over a curve in the lower half
plane, i.e. 
\begin{eqnarray*}
&&\int_0^\infty d\omega ^{\prime }e^{-i\omega ^{\prime }t}|\omega ^{\prime
+}\rangle \langle \omega ^{\prime +}| \\
&=&\int_0^\infty d\omega ^{\prime }e^{-i\omega ^{\prime }t}S(\omega ^{\prime
})|\omega ^{\prime -}\rangle \langle \omega ^{\prime +}|\rightarrow \\
&\rightarrow &e^{-iz_0t}|f_0\rangle \langle \widetilde{f_0}|+\int_0^{-\infty
}d\omega ^{\prime }e^{-i\omega ^{\prime }t}|(\omega ^{\prime
}-i0)^{+}\rangle \langle (\omega ^{\prime }-i0)^{+}|
\end{eqnarray*}
\begin{equation}
|f_0\rangle =[-2\pi i(ResS)_{z_0}]^{\frac 12}|z_0^{-}\rangle ,\quad \langle 
\widetilde{f_0}|=[-2\pi i(ResS)_{z_0}]^{\frac 12}\langle z_0^{+}|.
\label{4.15}
\end{equation}

Replacing (\ref{4.14}) and (\ref{4.15}) in (\ref{4.9}) we obtain 
\begin{eqnarray}
(\rho _t|O) &=&\int_0^\infty d\omega (\rho _0||\omega ^{+}\rangle \langle
\omega ^{+}|)O_\omega +  \nonumber \\
&&+e^{i(z_0^{*}-z_0)t}(\rho _0||\widetilde{f_0}\rangle \langle
f_0|O_{fluc}|f_0\rangle \langle \widetilde{f_0}|)+  \nonumber \\
&&+\int_0^{-\infty }d\omega ^{\prime }e^{i(z_0^{*}-\omega ^{\prime })t}(\rho
_0||\widetilde{f_0}\rangle \langle f_0|O_{fluc}|(\omega ^{\prime
}-i0)^{+}\rangle \langle (\omega ^{\prime }-i0)^{+}|)+  \nonumber \\
&&+\int_0^{-\infty }d\omega e^{i(\omega -z_0)t}(\rho _0||(\omega
+i0)^{+}\rangle \langle (\omega +i0)^{+}|O_{fluc}|f_0\rangle \langle 
\widetilde{f_0}|)+  \nonumber \\
&&+\int_0^{-\infty }d\omega \int_0^{-\infty }d\omega ^{\prime }e^{i(\omega
-\omega ^{\prime })t}\times  \nonumber \\
&&\times (\rho _0||(\omega +i0)^{+}\rangle \langle (\omega
+i0)^{+}|O_{fluc}|(\omega ^{\prime }-i0)^{+}\rangle \langle (\omega ^{\prime
}-i0)^{+}|)  \label{4.16}
\end{eqnarray}

The changes indicated in equations (\ref{4.14}) and (\ref{4.15}) are
possible if we impose on the states and observables the condition that $\rho
_{\omega \omega ^{\prime }}^{*}$ and $O_{\omega \omega ^{\prime }}$ have
well defined analytic extensions to the upper (lower) half plane in the
variable $\omega $ ($\omega ^{\prime }$)\footnote{%
Equations (\ref{4.14}) and (\ref{4.15}) requires the vanishing of the
integrals over a very big semicircle in the upper (lower) half plane in the
variable $\omega $ ($\omega ^{\prime }$). The presence of the exponential
factor $e^{iwt}$ ($e^{-i\omega ^{\prime }t}$), with $t>0$, makes the
requirement easier to be satisfied.}. In this case it is possible to prove
that $(O_{fluc})_{\omega \omega ^{\prime }}$ has also a well defined
analytic extension to the upper (lower) half plane in the variable $\omega $
($\omega ^{\prime }$). Expression (\ref{4.16}) correspond to the following
complex spectral decomposition of the identity (${\Bbb I}^{\dagger }$) and
Liouville-Von Neumann (${\Bbb L}^{\dagger }$) superoperators 
\begin{eqnarray}
{\Bbb I}^{\dagger } &=&\int_0^\infty d\omega |\Psi _\omega )(\widetilde{\Psi 
}_\omega |+|\Psi _{00})(\widetilde{\Psi }_{00}|+\int_0^{-\infty }d\omega
^{\prime }|\Psi _{0\omega ^{\prime }})(\widetilde{\Psi }_{0\omega ^{\prime
}}|+  \nonumber \\
&&+\int_0^{-\infty }d\omega |\Psi _{\omega 0})(\widetilde{\Psi }_{\omega
0}|+\int_0^{-\infty }d\omega \int_0^{-\infty }d\omega ^{\prime }|\Psi
_{\omega \omega ^{\prime }})(\widetilde{\Psi }_{\omega \omega ^{\prime }}|, 
\nonumber \\
{\Bbb L}^{\dagger } &=&(z_0^{*}-z_0)|\Psi _{00})(\widetilde{\Psi }%
_{00}|+\int_0^{-\infty }d\omega ^{\prime }(z_0^{*}-\omega ^{\prime })|\Psi
_{0\omega ^{\prime }})(\widetilde{\Psi }_{0\omega ^{\prime }}|+  \nonumber \\
&&+\int_0^{-\infty }d\omega (\omega -z_0)|\Psi _{\omega 0})(\widetilde{\Psi }%
_{\omega 0}|+  \nonumber \\
&&+\int_0^{-\infty }d\omega \int_0^{-\infty }d\omega ^{\prime }(\omega
-\omega ^{\prime })|\Psi _{\omega \omega ^{\prime }})(\widetilde{\Psi }%
_{\omega \omega ^{\prime }}|,  \label{4.17}
\end{eqnarray}
where 
\begin{eqnarray}
|\Psi _\omega ) &=&||\omega ^{+}\rangle \langle \omega ^{+}|)  \nonumber \\
(\widetilde{\Psi }_\omega | &=&(\omega |  \nonumber \\
|\Psi _{00}) &=&||\widetilde{f_0}\rangle \langle \widetilde{f_0}|)  \nonumber
\\
(\widetilde{\Psi }_{00}| &=&[2\pi i(ResS^{*})_{z_0^{*}}]^{-\frac 12}[-2\pi
i(ResS)_{z_0}]^{-\frac 12}(2\pi i\,Res_{\omega =z_0^{*}})(-2\pi
i\,Res_{\omega ^{\prime }=z_0})(\widetilde{\Phi }_{\omega \omega ^{\prime }}|
\nonumber \\
|\Psi _{0\omega ^{\prime }}) &=&||\widetilde{f_0}\rangle \langle (\omega
^{\prime }-i0)^{+}|)  \nonumber \\
(\widetilde{\Psi }_{0\omega ^{\prime }}| &=&[2\pi i(ResS^{*})_{z_0^{*}}]^{-%
\frac 12}(2\pi i\,Res_{\omega =z_0^{*}})(\widetilde{\Phi }_{\omega \omega
^{\prime }}|  \nonumber \\
|\Psi _{\omega 0}) &=&||(\omega +i0)^{+}\rangle \langle \widetilde{f_0}|) 
\nonumber \\
(\widetilde{\Psi }_{\omega 0}| &=&[-2\pi i(ResS)_{z_0}]^{-\frac 12}(-2\pi
i\,Res_{\omega ^{\prime }=z_0})(\widetilde{\Phi }_{\omega \omega ^{\prime }}|
\nonumber \\
|\Psi _{\omega \omega ^{\prime }}) &=&||(\omega +i0)^{+}\rangle \langle
(\omega ^{\prime }-i0)^{+}|)  \nonumber \\
(\widetilde{\Psi }_{\omega \omega ^{\prime }}| &=&(|(\omega +i0)^{+}\rangle
\langle (\omega ^{\prime }-i0)^{+}||  \label{4.18}
\end{eqnarray}

From these equations it is easy to prove that $(\widetilde{\Psi }_{00}|$, $(%
\widetilde{\Psi }_{0\omega ^{\prime }}|$, $(\widetilde{\Psi }_{\omega 0}|$
and $(\widetilde{\Psi }_{\omega \omega ^{\prime }}|$ have no energy nor
''trace'', i.e. 
\begin{eqnarray*}
(\widetilde{\Psi }_{00}|H) &=&(\widetilde{\Psi }_{0\omega ^{\prime }}|H)=(%
\widetilde{\Psi }_{\omega 0}|H)=(\widetilde{\Psi }_{\omega \omega ^{\prime
}}|H)=0 \\
(\widetilde{\Psi }_{00}|I) &=&(\widetilde{\Psi }_{0\omega ^{\prime }}|I)=(%
\widetilde{\Psi }_{\omega 0}|I)=(\widetilde{\Psi }_{\omega \omega ^{\prime
}}|I)=0,
\end{eqnarray*}
and therefore these generalized states cannot have an independent physical
meaning.

\section{Conclusions.}

For a quantum scattering problem, the continuous spectrum requires the
existence of well defined values of observables with diagonal singularities.
The Hamiltonian of the system, in momentum representation, is an example of
this class of observables.

Defining the states as functionals acting on the operators representing the
observables, more general states are allowed in the formalism. These new
class of states are not representable by wave functions nor by trace class
density operators.

The main conclusion of this paper is that the functional approach can give a
more complete description of the process. We proved that, even for initial
conditions representable by wave functions, the ''final'' state ($t=\infty $%
) is a well defined diagonal functional ($\rho _\infty ={\Bbb P}\rho _\infty 
$), which is a mixture of generalized eigenvectors of the free Hamiltonian $%
H_0$: 
\begin{equation}
\rho _\infty =\lim_{t\rightarrow \infty }\rho _t=\int d\overline{k}\,(\rho
_o||\overline{k}^{+}\rangle \langle \overline{k}^{+}|)(\overline{k}|,
\label{5.1}
\end{equation}
\[
(\overline{k}|\overline{k}^{\prime })=\delta (\overline{k}-\overline{k}%
^{\prime }),\quad |\overline{k})=|\overline{k}\rangle \langle \overline{k}%
|,\quad H_0|\overline{k}\rangle =|\overline{k}\rangle . 
\]
This state cannot be represented by a wave function or by a trace class
operator.

The final state depend on the initial condition, although different initial
conditions may converge into the same final state.

The time inversion of a state is well defined in the formalism 
\begin{eqnarray*}
\rho &=&\int d\overline{p}\,\rho _{\overline{p}}^{*}(\overline{p}|+\int \int
d\overline{p}\,d\overline{p}^{\prime }\,\rho _{\overline{p}\,\overline{p}%
^{\prime }}^{*}(\overline{p}\,\overline{p}^{\prime }|, \\
{\Bbb T}\rho &=&\int d\overline{p}\,\rho _{-\overline{p}}(\overline{p}|+\int
\int d\overline{p}\,d\overline{p}^{\prime }\,\rho _{-\overline{p}\,-\,%
\overline{p}^{\prime }}(\overline{p}\,\overline{p}^{\prime }|,
\end{eqnarray*}
and a ''weak intrinsic irreversibility'' appears: as

\[
{\Bbb T}\rho _\infty =\int d\overline{k}\,(\rho _o||-\overline{k}^{+}\rangle
\langle -\overline{k}^{+}|)^{*}(\overline{k}|, 
\]
and 
\[
{\Bbb U}_t{\Bbb T}\rho _\infty =e^{-i{\Bbb L}t}{\Bbb T}\rho _\infty ={\Bbb T}%
\rho _\infty , 
\]
the time evolution of the time inverted 'final' state cannot reproduce the
initial state. But this irreversibility appears for processes involving an
infinite amount of time, as the 'final' state is obtained with $t\rightarrow
\infty $. For a very big time $t_o<\infty $, the time inversion is possible
in principle, although it may be very difficult to prepare the state ${\Bbb T%
}\rho _{t_o}$ in practice.

It is interesting to emphasize that it is not necessary to consider the
analytic extensions of the ''components'' of the states to obtain these
results. Only regularity of the functions representing the off diagonal
parts of states and observables is required to expand the observables in
terms of the Lipmann-Schwinger eigenvectors, and to use the Riemann-Lebesgue
theorem for the deduction of (\ref{5.1}).

If, in addition, we assume that the ''components'' of states and observables
have well defined analytic extensions, as it is the case if Hardy class
functions are involved, a complex spectral decomposition of the time
evolution is possible. In section 4, we constructed a complex spectral
decomposition including ''generalized Gamov states'' related to the poles of
the ''$S$ matrix''. However, we proved that these generalized states have
zero energy and zero ''trace''. This is an expected result, consistent with
energy and probability conservation, because the ''Gamov states'' expand the
time dependent part of the physical states, which goes to zero for infinite
time. Therefore, in this formalism, the ''Gamov states'' cannot exist as
autonomous states, but only in a linear combination which should include the
time independent component (an eigenvector of the Lioville-Von Neumann
superoperator with zero eigenvalue).

\end{document}